\documentclass[twocolumn,aps,superscriptaddress,showpacs,nofootinbib,floatfix]{revtex4}
\usepackage{listings}
\usepackage[utf8]{inputenc}
\usepackage{hyperref}
\usepackage{amsmath}
\usepackage{amssymb}
\usepackage{amsfonts}
\usepackage{tabularx}
\usepackage{booktabs}
\usepackage{graphicx}
\usepackage{color}
\usepackage{multirow}
\usepackage[inline]{enumitem}
\usepackage[T2A,T1]{fontenc}
\graphicspath{{fig/}}
\definecolor{theblue}{RGB}{0,50,230}

\usepackage{hyperref}
\hypersetup{
  colorlinks=true,
  linkcolor=theblue,
  citecolor=theblue,
  urlcolor=theblue
}

\newcommand {\avg}[1]{\ensuremath{\langle\kern-1.0pt\langle#1\rangle\kern-1.0pt\rangle}}

\newlength\cmsFigWidth

\usepackage[normalem]{ulem}  % \sout{old text} for strikeout

\renewcommand\sout{\bgroup \color{red} \ULdepth=-.5ex \ULset}
\begin{document}

%%%%%%%%%%%%%%%%%%%%% Title %%%%%%%%%%%%%%%%%%%%%%

\title{Dynamical Pair Production at Sub-Barrier Energies for Light Nuclei}

%%%%%%%%%%%%%%%%%%%% Authors %%%%%%%%%%%%%%%%%%%%%
\author{T. Settlemyre}
\affiliation{Cyclotron Institute, Texas A\&M University, College Station, Texas 77843, USA}
\author{H. Zheng}
\affiliation{School of Physics and Information Technology, Shaanxi Normal University, Xi'an 710119, China}
\author{A. Bonasera}
\affiliation{Cyclotron Institute, Texas A\&M University, College Station, Texas 77843, USA}
\affiliation{Laboratori Nazionali del Sud, INFN, via Santa Sofia, 62, 95123 Catania, Italy}

%%%%%%%%%%%%%%%%%%%% Abstract %%%%%%%%%%%%%%%%%%%%%

\begin{abstract}
In the collision of two heavy ions, the strong repulsion coming from the Coulomb field is enough to produce $e^+e^-$ pair(s) from vacuum fluctuations.  The energy is provided by the kinetic energy of the ions and the Coulomb interaction at the production point.  If, for instance, the electron is located at the center of mass (C.M.) of the two ions moving along the \emph{z}-axis, and the positron is at a distance \emph{x} from the electron, the ions can be accelerated towards each other since the Coulomb barrier is lowered by the presence of the electron.  This screening results in an increase in the kinetic energy of the colliding ions and may result in an increase in the fusion probability of light ions above the adiabatic limit.
\end{abstract}

%\pacs{}
%\keywords{}
\maketitle

\section{introduction}
The exchange of a virtual photon is responsible for the Coulomb force. In this process virtual electron-positron pairs can be created and annihilated. These virtual charges  polarize the vacuum, resulting in a correction to the $1/r$ potential. Uehling was the first to derive the vacuum polarization correction to first order in the fine structure constant $\alpha$ \cite{Uehling:1935uj}, important in the analysis of $p-p$ scattering data \cite{prestonbook}. In  \cite{Settlemyre:2021tbe}, we showed that the vacuum polarization correction is of the order of one percent of the Coulomb energy in nuclear collision systems. This value seems small, but the strong fields in fission processes can be of the order of 200~MeV. A correction of the order of 2 MeV could noticeably affect the height of the Coulomb barrier, where the nuclear and Coulomb energies roughly cancel. A lower/higher Coulomb barrier increases/decreases the cross-section of sub-barrier fusion. In particular, carbon-carbon fusion in the cores of stars has been studied extensively, both theoretically~\cite{bonasera:fusion1, diaz, beck, esbensen, godbey} and experimentally \cite{aguilera, spillane, patterson, jiang,tumino, fruet}.

The vacuum polarization is not just a perturbative effect; production of real $e^+e^-$ pairs can occur during the dynamics in the presence of strong fields, when the available energy exceeds twice the electron mass \cite{Schwinger:1951nm, wongbook, Blaschke:2019pnj, carrollbook,Voskresensky:2021okp}. In this paper, we discuss the non-perturbative calculation of pair production for light nuclei. We show that in opportune conditions, the pair may screen the Coulomb repulsion between the ions giving them an extra acceleration towards each other.  This effect may increase the fusion cross-section above the adiabatic limit \cite{Kimura:2004dw, Krauss:1987onj, Engstler:1988tfw, Shoppa:1993zz, Musumarra:2001xd, Kimura:2005tq}.

\section{Schwinger Mechanism}
\begin{figure}[h]
\centerline{
\includegraphics[width=8.5cm]{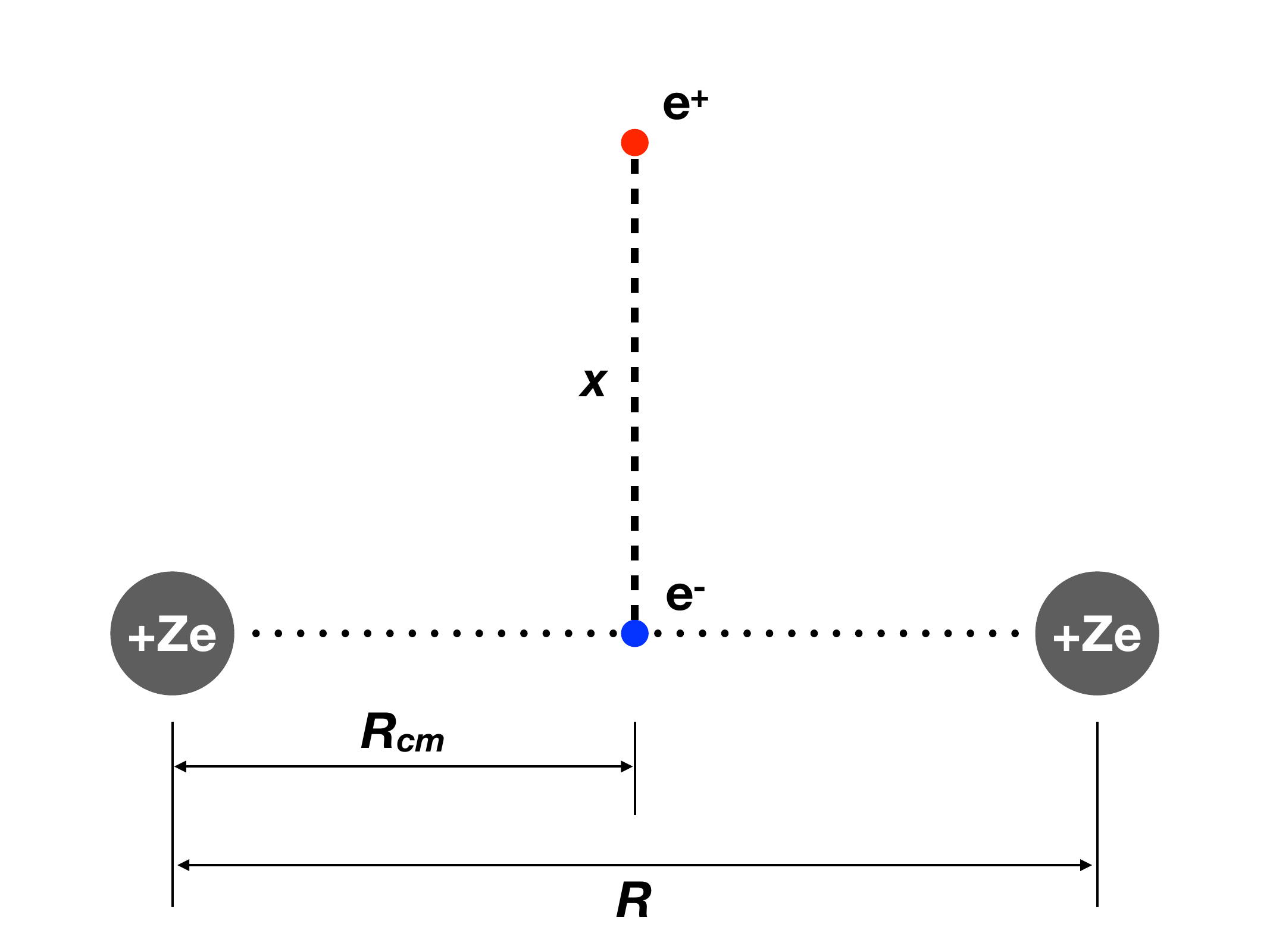}}
\caption{(Color online) Geometry of the pair production.}
\label{fig1}
\end{figure}

For the positron to become a real particle, it must tunnel from the vacuum through the Coulomb barrier and leave the electron behind. We compute the probability of tunneling through this barrier. We have two nuclei, each with charge $+Ze$ (for simplicity), a distance $R$ apart (Figure \ref{fig1}).  We assume that the electron is at the center of mass of the two nuclei and the positron is tunneling on a line perpendicular to the beam axis. Notice that the pair could be placed anywhere and the corresponding probabilities can be easily calculated.  Their values will be smaller than the chosen geometry. The distance from the electron to the positron is labeled by the coordinate $x$. The Coulomb energy of the positron is (in units where $4\pi\varepsilon_0$ = 1)
\begin{equation}
    V_+(R, x) = \frac{2Ze^2}{\sqrt{ \left( \frac{R}{2} \right)^2+ x^2}} - S(x) \frac{e^2}{x}, \label{vc}
\end{equation}
where $S(x)$ is a screening factor to be discussed in the sequel.  When the positron emerges from the barrier, it can have a momentum $p_T$ perpendicular to the $x$-axis, and the electron will have momentum $-p_T$.  To a good approximation \cite{Schwinger:1951nm, wongbook}, the positron satisfies the Klein--Gordon (K.G.) equation with energy
$E_+$,
\begin{equation}
    \left[(E_+-V_+(R, x))^2-p_x^2-m_T^2\right]\psi = 0, \label{KG}
\end{equation}
where $m_T = \sqrt{m^2+p_T^2}$ is the transverse mass of the positron.
The Dirac equation leads to the K.G. equation with an extra term ${\boldsymbol\alpha} \cdot \nabla V$, which comes from the spinor nature of the fermion wave function and gives only high-order effects in the tunneling probability; thus, it is neglected in this paper \cite{wongbook}. Following Wong \cite{wongbook}, we divide Equation (\ref{KG}) by $-2m_T$ to obtain
\begin{equation}
    \left[\frac{p_x^2}{2m_T} + \frac{m_T}{2} - \frac{(E_+ - V_+(R, x))^2}{2m_T}\right]\psi = 0.
\end{equation}

We have implicitly factored out the transverse plane wave part of the wave function. This is formally equivalent to the (time-independent) Schr\"odinger equation

\begin{equation}
     \left(\frac{p_x^2}{2m_T} + V_{eff} \right)\psi = E_{eff} \psi, \label{sch}
\end{equation}
for a particle of mass $m_T$ in an effective potential of
\begin{equation}
    V_{eff}(x) = \frac{m_T}{2} - \frac{(E_+ - V_+(R, x))^2}{2m_T}, \label{veff}
\end{equation}
with energy $E_{eff}=0$. The classical turning points are where $V_{eff}=0$, or
\begin{equation}
    E_+ - V_+(R, x) = \pm m_T. \label{tp}
\end{equation}
The maximum occurs when $E_+ = V_+(R, x)$ and the barrier height is given by $m_T/2$.

Considering a positron in an occupied negative energy state at $x = 0$, by definition, it has energy $E_+ \le V_+(R, 0)-m_T$. In order to tunnel into the positive energy region, its energy must be $E_+ \ge m_T$. Taking these conditions together, we obtain a constraint on $V_+$ for pair production, namely,
\begin{equation}
    V_+(R, 0) \geq 2m_T.  \label{vp2mt}
\end{equation}

This result confirms the intuition that pair production is possible for electrostatic energies exceeding twice the mass of the electron.

\section{Cross Section of Pair Production}
The total energy before $e^+e^-$ production is

\begin{equation}
    E_{cm} = E_k + V_{II}(R) =\frac{P^2}{2\mu} + \frac{Z^2e^2}{R}, \label{energy_before}
\end{equation}
where $\mu$ is the reduced mass of the colliding ions and we assume $R \ge R_1 + R_2$, the nuclear radii, i.e., beam energies below the Coulomb barrier. The $e^+e^-$ are produced with transverse mass $m_T$  at a relative distance $x_e$, which will be discussed below. The potential energy seen by the positron is
\begin{equation}
 V_+(R,x_e) = \frac{2Ze^2}{\sqrt{ \left( \frac{R}{2} \right)^2 + x_e^2 }} - S(x_e) \frac{e^2}{x_e}.
\end{equation}

The energy of the positron is the sum of its mass, kinetic energy, and potential energy,
\begin{equation}
E_+ = m_T + V_+(R,x_e).
\end{equation}

We have introduced a dynamical screening factor $S(x) = 1- \exp(-x/x_s)$. The choice $x_s=\frac{e^2}{2m_T}$, sometimes called the classical screening value, implies that, for $x\rightarrow 0$, the $e^+e^-$ are on top of each other and the mass is given by the Coulomb screened interaction, $S(x)e^2/x \rightarrow 2m_T$, which is the energy needed from an external source (the Coulomb field of the ions) to produce the pair, see Equation (\ref{vp2mt}). This assumption ensures energy conservation, avoiding the ultraviolet divergence of the Coulomb field.  This screening could come from the virtual particles in the vacuum. For instance, we imagine the vacuum as containing a density of pairs proportional to the energy density of the Coulomb field, which, in our units,~is
\begin{equation}
u=\frac{1}{8\pi}{\mathcal E}^2.
\end{equation}

A typical value for the electric field in our system is
\begin{equation}
{\mathcal E}=\frac{Ze}{(R/2)^2}.
\end{equation}

We divide the energy density $u$ by $2m_T$ to get the number density of $e^+e^-$ pairs
\begin{equation}
n_p= \frac{Z^2e^2}{\pi R^4 m_T} =\frac{2Z^2x_s}{\pi R^4}.
\end{equation}

The pairs we consider in our model originate in the region between the two nuclei, which we model as a cylinder of radius $x_s$ and length $R$. Multiplying the volume of this cylinder by the density of pairs obtained previously, we obtain the expected number of virtual pairs available to tunnel
\begin{equation}
    N_p = \frac{2 Z^2 x_s^3}{ R^3}. \label{pair_number}
\end{equation}

For two uranium nuclei with their surfaces touching, $N_p\approx 14$; for carbon in the same configuration, $N_p \approx 1.2$. This low value for light ions barely justifies a perturbative treatment of the production.

Since the total energy must be conserved, after production, we have:
\begin{equation}
    E_{cm} = E_k' + V_{II}(R) + V_+ (R,x_e) - \frac{4Ze^2}{R} + 2m_T. \label{energy_after}
\end{equation}

The system is completely symmetric, but a small fluctuation will push the $e^+$ away from the $e^-$ due to the Coulomb repulsion between the positron and the ions. The positron tunnels through the Coulomb barrier and exits at $x_e$ where its (and the electron’s) momentum along the $x$-direction is zero (Figure \ref{fig1}). At $x_e$, the total energy is given by Equation (\ref{energy_after}). If the positron is very fast compared to the ion motion, then we can assume the ions do not move much.

A microscopic calculation is needed to determine the final energy distribution between the electron and the positron. Our approximation is good if $m_T$  is large so that the pair has a good amount of kinetic energy when it is created. Notice that, in the case of very large $m_T$, the $e^+$ and $e^-$ emerge at about 180$^{\circ}$ in the center of the mass frame. Comparing our various expressions for the energy, Equations (\ref{energy_before}) and (\ref{energy_after}), we find the kinetic energy of the ions changes by an amount
\begin{equation}
   E_k' - E_k = -\left(V_+(R,x_e) - \frac{4Ze^2}{R} + 2m_T \right).  \label{kechange}
\end{equation}

Since $V_+(R,x_e) = E_+ - m_T$, we can also rewrite this as
\begin{equation}
 \Delta E_k=E_k' -E_k=-\left(E_++m_T-\frac{4Ze^2}{R}\right). \label{kechange1}
\end{equation}

And
\begin{equation}
E_+=\frac{4Ze^2}{R}-m_T-\Delta E_k\ge m_T. \label{eplus1}
\end{equation}

The last condition gives
\begin{equation}
R\le \frac{4Ze^2}{2m_T+\Delta E_k}, \label{rleq}
\end{equation}
that is the largest distance for which the production may occur. The condition $E_+ \le V_+(R, 0)-m_T$ implies $\Delta E_k \ge 2m_T$. We stress again that other pair configurations  are, of course, possible, for instance, by exchanging the positron and the electron in Figure \ref{fig1}.  Different configurations cost more energy and are less probable, but calculations can be easily performed for any configuration.

For illustration, we enforce the condition $V_{eff} = 0$ at $x = 0$.  There can be two solutions corresponding to
\begin{equation}
    E_+ = \frac{4Ze^2}{R} - 2m_T \pm m_T. \label{eplus}
\end{equation}
Thus, according to Equations (\ref{kechange1}) and (\ref{eplus1}), the ions either {\it gain} $2m_T$ of kinetic energy, or there is no change in kinetic energy at the moment of production. This situation is very interesting, especially in the sub-barrier fusion of light nuclei since, even in the case of zero kinetic energy gain from the ions, the presence of the electron in the middle of the two ions lowers the Coulomb barrier, thus enhancing the fusion probability \cite{Kimura:2004dw, Kimura:2005tq}.  We are interested in unbound positrons with $E_+>m_T$. This requirement, together with Equation (\ref{eplus}), gives a maximum transverse mass for dynamical pair production
\begin{equation}
m_{T, max}=\frac{2Ze^2}{R}.  \label{mtmax}
\end{equation}

Since our model only includes the Coulomb force between the ions, we only consider $R > R_1 +R_2$, where the nuclear force is not as important. For two $^{12}$C nuclei with their surfaces touching, Equation (\ref{mtmax}) gives a maximum transverse mass of 3.14 MeV. For $^{238}$U in the same condition, $m_{T,max}$ = 17.8 MeV. 
The corresponding effective potential for the two solutions is
\begin{equation}
    V_{eff}^{(1)}(R,x) =  \frac{m_T}{2} - \frac{[m_T + V_+ (R, x) - \frac{4Ze^2}{R} ]^2}{2m_T}, 
    \end{equation}
\begin{equation}
    V_{eff}^{(2)}(R,x) =  \frac{m_T}{2} - \frac{[3m_T + V_+ (R, x) - \frac{4Ze^2}{R} ]^2}{2m_T}.
\end{equation}

\begin{figure}[h]
\centerline{
\includegraphics[width=8.5cm]{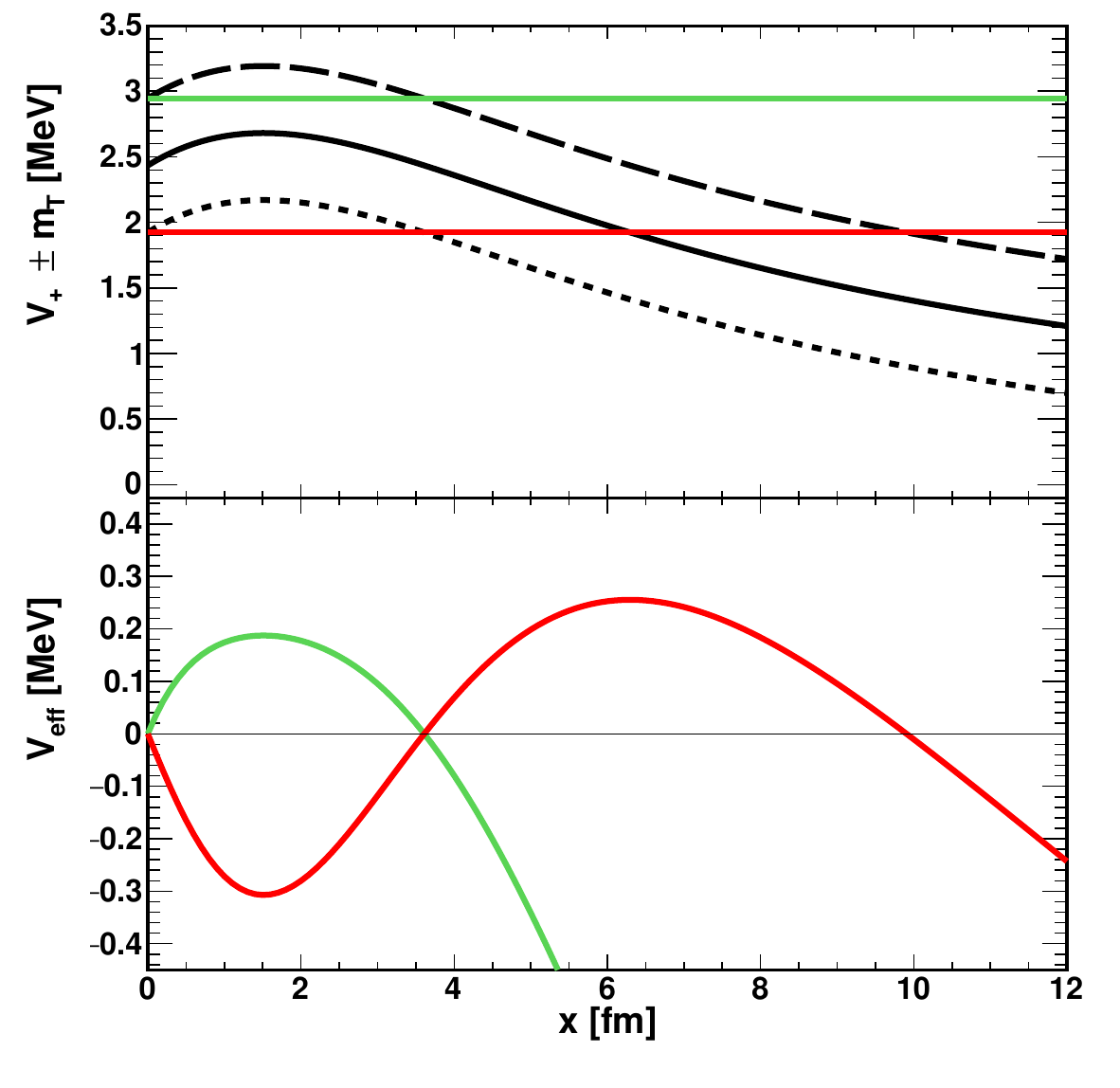}}
\caption{(Color online) An illustrative example when $V_{eff}(R,0)=0$. In the bottom panel we plot $V_{eff}$ vs. $x$ and the corresponding potential with full line ($\pm m_T=m_e$$-$top panel, dashed and dotted lines) seen by the positron. The calculations are performed for $^{12}$C+$^{12}$C collisions.}
\label{fig2}
\end{figure}

In Figure \ref{fig2}, we plot the effective potential (bottom panel) and the potential ($\pm m_T$$-$top panel) vs. the relative distance between the pair for the case discussed above.  The only acceptable solution is the lowest one given by the red line.  A simple inspection of the top panel shows that the positron for this case is initially in the negative energy region and tunnels to the positive one.  The other solution gives the positron already in the positive energy region (green line); thus, is not allowed by our proposed mechanism.  Other possible solutions can be found if $V_{eff}(R,x=0)<0$.  From this discussion, we learned that the two ions can {\it gain} kinetic energy because of the location of the electron (in the middle) and the positron (away from the ions) (Figure \ref{fig1}), and may enhance the sub-barrier fusion probability.

\begin{figure}[h]
\centerline{
\includegraphics[width=8.5cm]{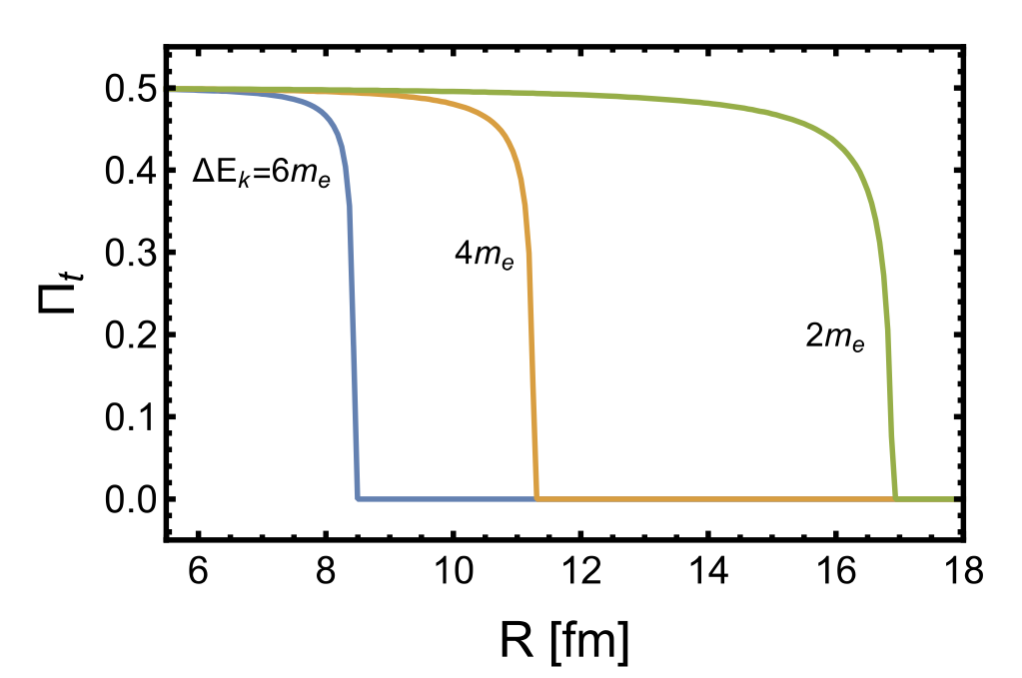}}
\caption{(Color online) Tunneling probability for the positron as a function of the relative distance of the two C ions and $E_{c.m.}$= 9.4 MeV and different values of $\Delta E_k$.}
\label{fig3}
\end{figure}

The tunneling probability for the positron is given by:
\begin{equation}
    \Pi_t = [1+\exp(2A)]^{-1},
\end{equation}
where $A$ is the imaginary action integrated between the turning points of the effective potential (see for instance Figure \ref{fig2}-bottom).  The action can be calculated numerically; some case results are displayed in Figure \ref{fig3} with $m_T=m_e$. In the calculations, different values of $\Delta E_k=E_k'-E_k$ (see Equation (\ref{eplus1})) have been assumed.  The lowest value of $R$ is given by the classical turning point, i.e.,:
\begin{equation}
R_i=\frac{Z^2e^2}{E_{c.m.}}. \label{eqrm}
\end{equation}

Since $E_+\ge m_T$, we can easily estimate the value of R where the probability becomes zero:
\begin{equation}
R_x=\frac{4Ze^2}{\Delta E_k + 2m_T}, \label{eqrx}
\end{equation}
which is consistent with Equation (\ref{rleq}). It is easy to show that $R_x\ge R_i$ if:
\begin{equation}
E_{c.m.}\ge \frac{Z}{4}(\Delta E_k+2m_T).
\end{equation}
From these results and Figure \ref{fig3}, we can safely assume that $\Pi_t=0.5$ for $R_i<R<R_x$.

It should not be surprising that the probability is of the order of 0.5 since the maximum height of the barrier is $m_T/2$ (see Equation (\ref{veff})), and the barrier width is of the order of 10 fm. The probability goes to zero when $E_+\rightarrow m_T$ and $x\rightarrow \infty$.  This also agrees with our estimate of the number of pairs produced in the cylinder of radius $x_s$, which, for C+C, is of the order of one.  For heavier nuclei, this calculation must be performed for each distance and all the created pairs must be followed microscopically since barriers may be modified by the presence of previously created pairs and there may not be enough energy to produce another pair after the first one.  The probability cutoffs in the figure are essentially determined by energy conservation for each value of $\Delta E_k$.

Since this is a dynamical process, times are important.  A characteristic time for pair production is given by the Heisenberg principle:

\begin{equation}
\Delta \tau = \frac{\hbar}{2m_T}, \label{deltatau}
\end{equation}
thus, the rate at which a given virtual pair can attempt to tunnel is $\Delta\tau^{-1}$. 

There is a second characteristic time for the tunneling process.  A simple inspection of Figure \ref{fig2} (bottom), shows that the positron may be trapped by the Coulomb barrier up to the inner turning point.  This is analogous to the number of assaults per unit time in the theory of alpha decay, fission, etc. This quantity may be estimated by the ratio of the distance traveled by the positron before hitting the inner barrier (of the order of a few Fermis from Figure \ref{fig2}) divided by its average speed.  For transverse masses equal to the rest mass of the electron, the corresponding time is smaller than the time obtained from the Heisenberg uncertainty principle, and we will use the value given in Equation (\ref{deltatau}) for an estimate of the cross-section. Microscopic dynamical calculations are needed for heavier systems when more than one pair may be produced and energy conservation must be fulfilled.

Here, we use simple and transparent physical arguments to estimate the value of the cross-section for pair production.  We write the cross-section as:
\begin{equation}
\sigma(E_{c.m.})=\frac{\pi \hbar^2}{2\mu E_{c.m.}}\sum_{l=0}^n(2l+1)\Pi_lP_H.
\end{equation}

Since we are interested in sub-barrier reactions, we only consider the $l=0$ case  and we fix $\Pi_0=\Pi_t=0.5$, as discussed above, and $P_H= 1 - \exp(-\tau/\Delta \tau)$ (see Equation (\ref{deltatau})). Thus, in order to estimate the cross-section, we need the $\tau$ it takes for the ions to travel from $R_x$ to $R_i$ (Equations (\ref{eqrm}) and (\ref{eqrx})). For the case of the zero impact parameter, this can be computed exactly. With $m_T=m_e$, the cross-section is:
\begin{eqnarray}
\sigma_0(E_{c.m.})&=&\frac{\pi \hbar^2}{2\mu E_{c.m.}}0.5 \left(1 - \exp\left[-\frac{\tau}{\Delta \tau}\right]\right)\nonumber\\
\tau &=&\sqrt{2\mu} \bigg( \frac{R_x}{E_{cm}} \sqrt{E_{cm}-\frac{Z^2 e^2}{R_x}}\label{cross-section}\\
&& + \frac{Z^2 e^2}{E_{cm}^{3/2}} \text{arctanh}\sqrt{1 - \frac{Z^2 e^2}{R_x E_{cm}}} \bigg). \nonumber
\end{eqnarray}

\begin{figure}[h]
\centerline{
\includegraphics[width=9cm]{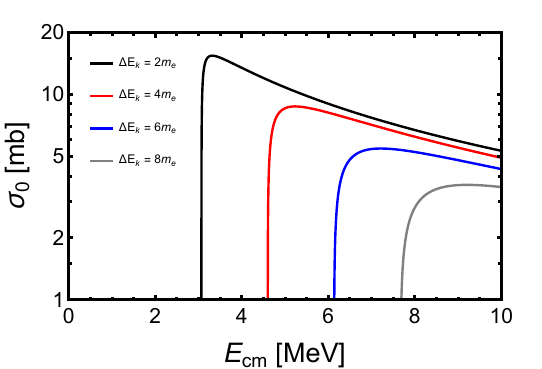}}
\caption{(Color online) Upper limit for the integrated cross-section for $e^+e^-$ production in $^{12}$C+$^{12}$C scattering below the Coulomb barrier for different values of $\Delta E_k$. We stress that $\Delta E_k \geq 0$.}
\label{fig4}
\end{figure}

Equation (\ref{cross-section}) gives a lower limit for $E_{cm}$. For $^{12}$C+$^{12}$C, we find the maximum number of pairs produced in the collisions by summing over the trajectory without taking into account the energy loss after a pair is produced. The maximum is attained near \linebreak $E_{c.m.}$= 4 MeV ($\Delta E_k=2m_e$) in Figure \ref{fig4}. Clearly, the maximum number of pairs produced in the collisions, and the relative cross-section of Figure \ref{fig4}, critically depends on the ultraviolet cutoff $x_s$ discussed above and it must be confirmed or modified by future experimental data.  Furthermore, microscopic calculations following the heavy ion trajectory and the dynamics of one or more pairs created during the time evolution must be implemented in order to make predictions for heavier colliding nuclei and collisions of different mass number nuclei.

\section{Summary}\label{summary}

In conclusion, we have discussed pair production from a vacuum within the Schwinger formalism.  We have shown the conditions for tunneling and the possibility that, if the electron is situated at the c.m. of the colliding nuclei, extra screening may occur.  This screening may enhance sub-barrier fusion of light nuclei above the adiabatic limit. For $^{12}$C+$^{12}$C collisions, we predict $E_{c.m.}\ge 1$ MeV for this effect to occur.  The cross-sections are of the order of mb or less.  These predictions call for detailed experimental investigation of pair production for this system, and also their energies,  in coincidence with fusion fragments to be able to extract correlation functions.  An enhancement may be shown by performing a correlation between fusion events with and without pair production.

\section*{Acknowledgements}
This research was funded in part by the United States Department of Energy under Grant \# DE-FG03-93ER40773 and the NNSA Grant No. DENA0003841 (CENTAUR) and by the National Natural Science Foundation of China (Grant Nos. 11905120 and 11947416).


\begin{thebibliography}{99}
\bibitem{Uehling:1935uj}
Uehling, E.A. Polarization effects in the positron theory. \emph{Phys. Rev.} \textbf{1935}, \emph{48}, 55--63.

\bibitem{prestonbook}Preston, M.; Bhaduri, R. \emph{Structure of the Nucleus}; Addison-Wesley Publishing Company: Reading, MA, USA, 1975.


\bibitem{Settlemyre:2021tbe}
Settlemyre, T.; Zheng, H.; Bonasera, A.
Coulomb field correction due to virtual e+ e\ensuremath{-} production in heavy ion collisions.
\emph{Nucl. Phys. A} \textbf{2021}, \emph{1015}, 122282.


\bibitem{bonasera:fusion1}
Bonasera, A.; Natowitz, J.B. Calculation of the $^{12}C^{12}C$ sub-barrier fusion cross section in an imaginary-time-dependent mean field theory. \emph{Phys. Rev. C} \textbf{2020}, \emph{102}, 061602.

\bibitem{diaz}
Diaz-Torres, A.; Wiescher, M. Characterizing the astrophysial $s$ factor for $^{12}C^{12}C$ fusion with wave-packet dynamics. \emph{Phys. Rev. C} \textbf{2018}, \emph{97}, 055802.

\bibitem{beck}
Beck, C.; Mukhamedzhanov, A.M.; Tang, X. Status on $^{12}C^{12}C$ fusion at deep subbarrier energies: impact of resonances on astrophysical $s^*$ factors. \emph{Eur. Phys. J. A} \textbf{2020}, \emph{56}, 87.

\bibitem{esbensen}
Esbensen, H.; Tang, X.; Jiang, C.L. Effects of mutual excitations in the fusion of carbon isotopes. \emph{Phys. Rev. C} \textbf{2011}, \emph{84}, 064613.

\bibitem{godbey}
Godbey, K.; Simenel, C.; Umar, A.S. Absence of hindrance in a microscopic $^{12}C^{12}C$ fusion study. \emph{Phys. Rev. C} \textbf{2019}, \emph{100}, 024619.

\bibitem{aguilera}
Aguilera, E.F.; Rosales, P.; Martinez, E.; Murillo, G.; Fernandez, M.; Berdejo, H.; Lizcano, D.; Camacho, A.G.; Policroniades, R.; Varela, A.; et al. New $\gamma$-ray measurements for $^{12}C^{12}C$ sub-coulomb fusion: toward data unification. \emph{Phys. Rev. C} \textbf{2006}, \emph{73}, 064601.

\bibitem{spillane}
Spillane, T.; Raiola, F.; Rolfs, C.; Schurmann, D.; Strieder, F.; Zeng, S.; Becker, H.-W.; Bordeanu, C.; Gialanella, L.; Romano, M.; et al. $^{12}C^{12}C$ fusion reactions near the Gamow energy. \emph{Phys. Rev. Lett.} \textbf{2007}, \emph{98}, 122501.

\bibitem{patterson}
Patterson, J.R.; Winkler, H.; Zaidins, C.S. Experimental investigation of the stellar nuclear reaction $^{12}C^{12}C$ at low energies. \emph{Astrophys. J.} \textbf{1969}, \emph{157}, 367.

\bibitem{jiang}
Jiang, C.L.; Santiago-Gonzalez, D.; Almarez-Calderon, S.; Rehm, K.E.; Back, B.B.; Auranen, K.; Avila, M.L.; Ayangeakaa, A.D.; Bottoni, S.; Carpenter, M.P.; et al. Reaction rate for carbon burning in massive stars. \emph{Phys. Rev. C} \textbf{2018}, \emph{97}, 012801.

\bibitem{tumino}
Tumino, A.; Spitaleri, C.; La Cognata, M.;  Cheurbini, S.; Guardo, G.L.; Gulino, M.; Hayakawa, S.; Indelicato, I.; Lamia, L.; Petrascu, H.; et al. An increase in the $^{12}C^{12}C$ fusion rate from resonances at astrophysical energies. \emph{Nature} \textbf{2018}, \emph{557}, 687--690.

\bibitem{fruet}
Fruet, G.; Courtin, S.; Heine, M.; Jenkins, D.G.; Adsley, P.; Brown, A.; Canavan, R.; Catford, W.N.; Charon, E.; Curien, D.; et al. Advances in the direct study of carbon burning in massive stars. \emph{Phys. Rev. Lett}. \textbf{2020}, \emph{124}, 192701.

\bibitem{Schwinger:1951nm}
Schwinger, J.S. On gauge invariance and vacuum polarization. \emph{Phys. Rev.} \textbf{1951}, \emph{82}, 664--679.

\bibitem{wongbook}
Wong, C. \emph{Introduction to High-Energy Heavy-Ion Collisions}; World Scientific Publishing: Singapore, 1994.


\bibitem{Blaschke:2019pnj}
Blaschke, D.B.; Juchnowski, L.; Otto, A. Kinetic Approach to Pair Production in Strong Fields--Two Lessons for Applications to Heavy-Ion Collisions. \emph{Particles} \textbf{2019}, \emph{2}, 166--179.

\bibitem{carrollbook}
Carroll, B.W.; Ostlie, D.A. \emph{An Introduction to Modern Astrophysics}, 2nd ed.; Addison-Wesley: San Francisco, CA, USA, 2007. 

\bibitem{Voskresensky:2021okp}
Voskresensky, D.N.
Electron-positron vacuum instability in strong electric fields. Relativistic semiclassical approach.
\emph{Universe} \textbf{2021}, \emph{7}, 104.

\bibitem{Kimura:2004dw}
Kimura, S.; Bonasera, A. Chaos Driven Fusion Enhancement Factor at Astrophysical Energies.
\emph{Phys. Rev. Lett.} \textbf{2004}, \emph{93}, 262502.

\bibitem{Krauss:1987onj}
Krauss, A.; Becker, H.W.; Trautvetter, H.P.; Rolfs, C. Astrophysical S (E) factor of 3 He (3 He, 2p) 4 He at solar energies. \emph{Nucl. Phys. A} \textbf{1987}, \emph{467}, 273--290.


\bibitem{Engstler:1988tfw}
Engstler, S.; Krauss, A.; Neldner, K.; Rolfs, C.; Schr\"oder, U.; Langanke, K.
Effects of electron screening on the 3 He (d, p) 4 He low-energy cross sections. \emph{Phys. Lett. B} \textbf{1988}, \emph{202}, 179--184.

\bibitem{Shoppa:1993zz}
Shoppa, T.D.; Koonin, S.E.; Langanke, K.; Seki, R. One- and two-electron atomic screening in fusion reactions. \emph{Phys. Rev. C} \textbf{1993}, \emph{48}, 837--840.

\bibitem{Musumarra:2001xd}
Musumarra, A.; Pizzone, R.G.; Blagus, S.; Bogovac, M.; Figuera, P.; Lattuada, M.; Milin, M.D.; Miljanic, M.; Pellegriti G.; Rendic, D.; et al.
Improved information on the 2H (Li-6, alpha) He-4 reaction extracted via the [trojan horse] method. \emph{Phys. Rev. C} \textbf{2001}, \emph{64}, 068801.
 
\bibitem{Kimura:2005tq}
Kimura, S.; Bonasera, A.; Cavallaro, S. Influence of chaos on the fusion enhancement by electron screening. \emph{Nucl. Phys. A} \textbf{2005}, \emph{759}, 229--244.

\end{thebibliography}
\end{document}